\documentclass{emulateapj}
\usepackage{apjfonts}
\pdfoutput=1
\makeatletter

\newcommand{\bc}{\begin{center}}
\newcommand{\ec}{\end{center}}
\newcommand{\be}{\begin{equation}}
\newcommand{\ee}{\end{equation}}
\newcommand{\ba}{\begin{eqnarray}}
\newcommand{\ea}{\end{eqnarray}}
\newcommand{\bt}{\begin{tabular}}
\newcommand{\et}{\end{tabular}}

\def\farcs{\hbox{$.\!\!^{\prime\prime}$}}

\def\farcs{\hbox{$.\!\!^{\prime\prime}$}}

\begin{document}

\title{Chandra observation of the
relativistic binary J1906+0746}

\author{O.\  Kargaltsev\altaffilmark{1} and   G.\ G.\
Pavlov\altaffilmark{2}}

\altaffiltext{1}{ Dept.\ of Astronomy, University of Florida, Bryant
Space Science Center, Gainesville, FL 32611; oyk100@astro.ufl.edu}

\altaffiltext{2}{The Pennsylvania State University, 525 Davey Lab,
University Park, PA 16802, USA; pavlov@astro.psu.edu}

\begin{abstract}

PSR J1906+0746
is a young radio
pulsar ($\tau = 112$ kyr, $P=144$ ms) in a tight binary ($P_{\rm orb}=3.98$ hr)
with a
 compact high-mass companion ($M_{\rm comp}\simeq 1.36 M_\odot$),
at the distance of about 5 kpc.
 We observed this unique
relativistic binary
 with the
{\sl Chandra} ACIS detector for 31.6 ks.
 Surprisingly, not a single photon was detected within the  $3''$ radius
  from the J1906+0746 radio position.  For a plausible
range of hydrogen column densities,
 $n_{\rm H}=(0.5$--$1)\times10^{22}$ cm$^{-2}$,
the nondetection corresponds to
  the  $90$\% upper limit of
(3--$5)\times10^{30}$ erg s$^{-1}$ on the unabsorbed
0.5--8 keV  luminosity
   for the power-law model with $\Gamma=1.0$--2.0,
and
     $\sim 10^{32}$ erg s$^{-1}$ on the bolometric luminosity of the thermal emission from the NS surface.
 The inferred limits are
  the lowest known for
  pulsars with spin-down
   properties similar to those of PSR J1906+0746.
    We have also tentatively detected a puzzling extended structure which
     looks like  a tilted
   ring with
   a radius of  $1.6'$ centered on the pulsar.
The measured 0.5--8 keV flux of the feature,
$\approx3.1\times10^{-14}$ erg cm$^{-2}$ s$^{-1}$,
implies an unabsorbed luminosity of
  $1.2\times10^{32}$ erg s$^{-1}$
($4.5\times10^{-4}$ of the pulsar's $\dot{E}$)
for
     $n_{\rm H}=0.7\times10^{22}$ cm$^{-2}$.
   Although
   all conventional interpretations of the ring appear to be problematic, the pulsar-wind nebula with an unusually underluminous pulsar remains the most
viable interpretation.

\end{abstract}


\section{Introduction}

{\sl Chandra} and {\sl XMM-Newton} observations of isolated young and middle-aged pulsars
  reveal X-ray emission from the neutrons star (NS) surfaces and
magnetospheres as well as extended
 emission from pulsar wind nebulae (PWNe; see Kargaltsev \& Pavlov 2008 for a recent review;
  KP08 hereafter).
 The analysis of the X-ray properties of
the pulsars and PWNe observed with {\sl Chandra}  shows
a large scatter in their X-ray efficiencies,
  $\eta_X=L_{X}/\dot{E}$
(e.g., $10^{-5}\lesssim \eta_X\lesssim 10^{-1}$ for the combined pulsar
and PWN emission; KP08; Li et al.\ 2008). Among the most underluminous
 are several young pulsars, $\tau\sim 30$--100 kyr,
some of which have not been detected (e.g., PSR J1913+1011, which has the lowest known X-ray efficiency,
 $\eta  < 8\times 10^{-6}d_{5}^2$; KP08).
 Although
some of the very low efficiencies could be due to an underestimated
 distance, the growing number of underluminous pulsars (Newman et al., in
preparation) and the very different efficiencies of
pulsars with well known distances
  suggest that the inaccurate distances can be only
 part of
 the  story.

    PSR J1906+0746 (hereafter PSR\,J1906) was discovered by
Lorimer et al.\ (2006) in the
     ALFA
	Arecibo Pulsar Survey (Cordes et al.\ 2006).
 PSR\,J1906 is a young, 144 ms pulsar
in a highly relativistic
 orbit ($P_{\rm orb}=3.98$ hr) with eccentricity $e=0.085$ and
gravitational wave coalescence time of  300 Myr.
The companion mass
of  $(1.36\pm0.02) M_{\odot}$
 (Kasian et al.\ 2007),
 the substantial
 eccentricity  of the binary orbit,
 the small
spin-down age ($\tau=P/2\dot{P}=112$ kyrs),
the significant spin-down energy loss rate ($\dot{E}=2.7\times 10^{35}$ erg
 s$^{-1}$),
  and
  the relatively high
   magnetic field ($B\approx1.7\times 10^{12}$ G)
 suggest that the pulsar could be a young, second-born NS  in the double NS binary (DNSB)  J1906+0746 (hereafter J1906). In this case
we would expect the companion to be
 a millisecond pulsar (similar to PSR J0737--3039A
in the double pulsar binary J0737--3039),
recycled during an earlier phase of accretion. However, despite deep
radio searches, the second pulsar in J1906 has not been detected,
perhaps
 because of  the unfavorable orientation of its radio beam
(Lorimer et al.\ 2006). Alternatively,
J1906 could be similar to
the NS binary J1141--6545, in which the  companion is  a
massive, $0.99\pm0.02 M_{\odot}$, white dwarf (WD) on an
eccentric orbit
 around the 1.45 Myr old pulsar\footnote{A somewhat older, $\tau\approx30$ Myrs, pulsar B2303+46 is also known to be in a binary with a factor of 60
wider eccentric orbit and $(1.3\pm0.1)  M_{\odot}$ WD companion (van Kerkwijk \& Kulkarni 1999).} (Kaspi et al.\ 2000; Bailes et al.\ 2003).
  A main sequence companion is ruled out by the binary evolution scenarios (see e.g., Stairs 2004, and references therein)
   unless the binary companions did not evolve together
 (i.e., the binary was formed recently through a gravitational capture).

 The recent discoveries of X-ray emission from the
tight pulsar binaries
J0737--3039 (McLaughlin et al.\ 2004; Campana et al.\
2004; Pellizzoni et al.\ 2004, 2008),
B1534+12 (Kargaltsev et al.\ 2006)
and B1957+20 (Stappers et al.\ 2003)
 have drawn
substantial interest to the high-energy emission  mechanisms in such
systems,
particularly
because
one can infer the properties
of pulsar winds
and binary companions via observations of X-ray emission
from intrabinary shocks.
 Although relatively distant, J1906
appears to be
 a good
 candidate for detecting X-ray emission from yet another
tight NS binary
because  PSR\,J1906
 is young and energetic (e.g., the $L_{X}$--$\dot{E}$ relation from
Possenti et al.\ 2002 predicts the nonthermal X-ray luminosity of
$\simeq10^{32}$ erg s$^{-1}$ in 2--10 keV). This prompted us to
carry out an exploratory {\sl Chandra} observation,
  results of which we  describe below.

\begin{figure}
 \centering
\includegraphics[width=3.3in,angle=0]{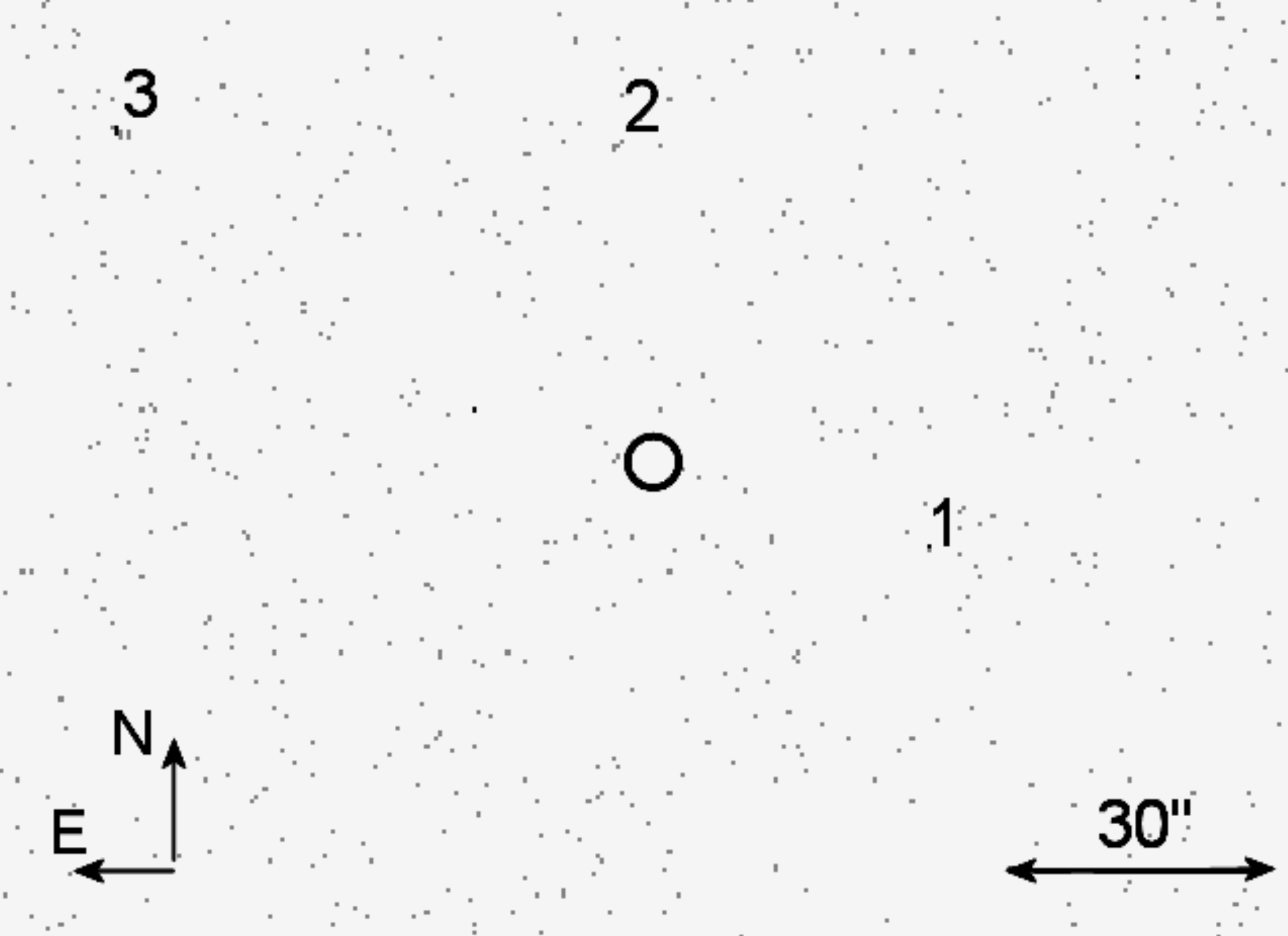}
\caption{
ACIS image of the J1906 vicinity (pixel size $0.492''$)
in the 0.5-8 keV band. The circle
($R=3''$) is centered on the radio position of PSR\,J1906  (Lorimer
et al.\ 2006). There are no X-ray counts within this circle.
Notice that the quoted uncertainty of the  radio position is much smaller than the circle shown (see text).  The three nearby sources
are numbered
 according to Table 1.}
\end{figure}

\section{Observations and data analysis}

We observed J1906 with the Advanced CCD Imaging Spectrometer (ACIS)
aboard {\sl Chandra} on 2007 August 10 (start time
  54,322.0374098  MJD).
 The useful scientific exposure time was 31,610 s. The
observation was carried out in Very Faint mode, and the pulsar was
imaged  $45.8''$ off-axis on the I3 chip of the ACIS-I array (the other
activated chips were I0, I1, I2 and S3). The detector was operated in
Full Frame mode, which provides a time resolution of 3.24 s.
The data were reduced using the {\sl Chandra} Interactive Analysis
of Observations (CIAO) software (ver.\ 4.0; CALDB ver.\ 3.4.0).
To maximize the signal-to-noise ratio, we choose the energy range
of 0.5--8 keV for the imaging and spectral analysis.

   \begin{table}
\caption[]{X-ray sources near the radio position of PSR\,J1906}
\begin{center}
\begin{tabular}{ccccccccc}
\tableline\tableline \#\tablenotemark{a} & $\Delta$\tablenotemark{b} & R.~A.   &
 Decl. & $N_{0.5-2}$\tablenotemark{c}    &  $N_{2-8}$\tablenotemark{d}    \\
\tableline
      1&  $33.2''$           &
        $19^{\rm h}06^{\rm m}46\fs529$ & $07^{\circ}46'18\farcs47$     &  4  &  0  \\
    2&     $36.7''$     &
      $19^{\rm h}06^{\rm m}48\fs954$ &  $07^{\circ}47'04\farcs84$     &  4    &  0  \\
     3&          $72.8''$    &
     $19^{\rm h}06^{\rm m}52\farcs846$ &  $07^{\circ}47'06\farcs47$     &  7    &  2           \\
     \tableline
\end{tabular}
\end{center}
\footnotesize{\tablenotetext{a}{Source number in  Figure 1.}
\tablenotetext{b}{Angular separation  from the radio pulsar position.}
\tablenotetext{c}{Number of counts in 0.5--2 keV band.}\tablenotetext{d}{Number of counts in 2--8 keV band.}
}
\end{table}

\subsection{X-ray data}

We found no X-ray source near
 the pulsar's radio position R.A.=$19^{\rm h}06^{\rm m}48\fs673(6)$,  Decl.=$07^{\circ}46'28\farcs6(3)$  reported by Lorimer et al.\
 (2006) from radio  timing.
  Figure 1 shows that there are no
  photons   in the  0.5--8 keV band within the $3''$ radius circle
around the radio position.
This radius is a factor of 8 larger than the $0.37''$
  uncertainty obtained by adding in quadratures the uncertainties
of the radio-timing
 position ($\sigma_{r}=0.31''$) and the {\sl Chandra} aspect
solution ($\sigma_{a}\approx 0.2''$; e.g., Pavlov et al.\ 2009,
 and references therein).
 Therefore, we conclude that
   J1906 is not detected in the {\sl Chandra} ACIS image.
   The  three nearest
    X-ray sources  (with $\geq4$ counts in the $r=1.5''$ aperture)
are offset by $33''$--$73''$
from the radio pulsar position (see Figure 1 and Table 1).

\begin{figure*}
\hspace{-0.3cm}
\includegraphics[width=7.3in,angle=0]{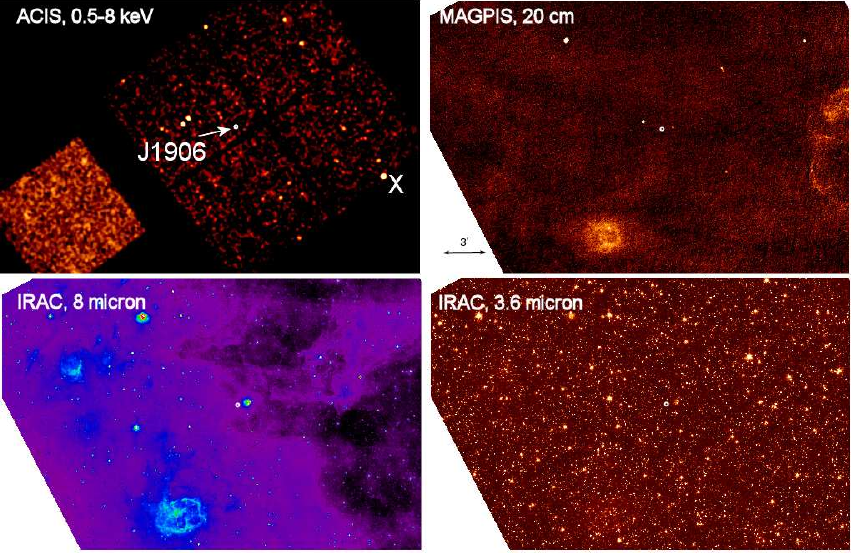}
\caption{ $18'\times29'$ images of the J1906 field at different wavelengths.
The radio position of J1906 is shown by the small circle.
 The ACIS image is binned to $2''$ pixels
 and smoothed with an $r=6''$ gaussian kernel. }
\end{figure*}

 We have used circular annuli centered on the pulsar radio position to measure the local background count rate out to $r=30''$ from the pulsar.
   We find the background
   to be rather uniform (the surface brightnesses measured from the individual annuli are consistent with a constant surface brightness within
   the uncertainties).
  The mean background
  surface brightness
is $0.035\pm0.003$ counts arcsec$^{-2}$.
   Thus, within the
 $r=3''$
 extraction aperture one would
expect
to detect about one count from the background.
 For a Poissonian distribution,
the nondetection with zero counts translates into the upper limit
$N<-\ln(1-{\rm CL})$ counts at the confidence level CL,
which gives the 90\% count rate upper limit of
 $7.3\times10^{-5}$
  counts s$^{-1}$.

  To constrain the absorption in the direction toward J1906, we  fit
the spectrum of the brightest X-ray source in the field (marked X in Fig.\ 2) located at the
 edge of the ACIS-I field-of-view  ($\approx 11'$ west of J1906 at
R.A.$=19^{\rm h}06^{\rm m}06\fs 6$ and Decl.$=07^{\circ}42'58''$).
  The spectrum (total 176 counts in the 0.5--8 keV band, of which 11\% are from the background) fits well ($\chi_{\nu}^2=0.74$ for $\nu=14$)
  an absorbed power-law (PL) model
with photon index
$\Gamma=1.3\pm0.2$ and
hydrogen column density $n_{\rm H,22}\equiv n_{\rm H}/10^{22}\,{\rm cm}^{-2}
=1.1\pm0.3$.
On the other hand,
   the PSR\,J1906's  dispersion
  measure DM = 218 pc cm$^{-3}$
   gives  $n_{\rm H,22}=0.65$,
assuming
     10\% ISM ionization. Finally,
   the total Galactic HI
   column  in the
    direction toward J1906  (galactic coordinates: $l=41.60^{\circ}$, $b=0.15^{\circ}$) is
$n_{\rm HI, 22} = 1.5$--1.9,
according to Dickey \& Lockman (1990), while
     Kalberla et al.\ (2005) give a slightly lower value,
$n_{\rm HI, 22} = 1.4$--1.6.
The distance to the edge of the Galaxy
in this direction is $\sim 15$ kpc,
 while the
distance to J1906,
based on the pulsar's
dispersion
measure, is about 5 kpc
(4.5 kpc and 5.4 kpc for the galactic electron density distributions
by Taylor \& Cordes 1993 and Cordes \& Lazio
   2002, respectively; we will scale all distance-dependent quantities to $d=5$ kpc below).
    Thus, a plausible range for the absorbing column toward J1906 is
$n_{\rm H, 22} \approx 0.5$--1.0.

  To calculate the upper limits on the X-rays fluxes from PSR\,J1906
   and its putative compact
   PWN, we
  consider two plausible spectral models: (1) a nonthermal
PL spectrum from the pulsar
magnetosphere and/or from a compact unresolved PWN, and
(2) a thermal spectrum
emitted from the NS surface.
    We then calculate the
     upper limits on the 0.5--8 keV unabsorbed flux as a function of
photon index for the PL model (Fig.\ 3) and on the bolometric luminosity
as a function of the NS surface temperature  for the
blackbody (BB) model (Fig.\ 4),
      using {\sl Chandra\/}
PIMMS\footnote{See http://cxc.harvard.edu/toolkit/pimms.jsp} for
$n_{\rm H,22}=0.5$, 0.7, and 1.0.
     The   implications of these results are discussed in \S\S\,3.1 and 3.2.

\begin{figure}
\hspace{-0.5cm}
\includegraphics[width=2.7in,angle=90]{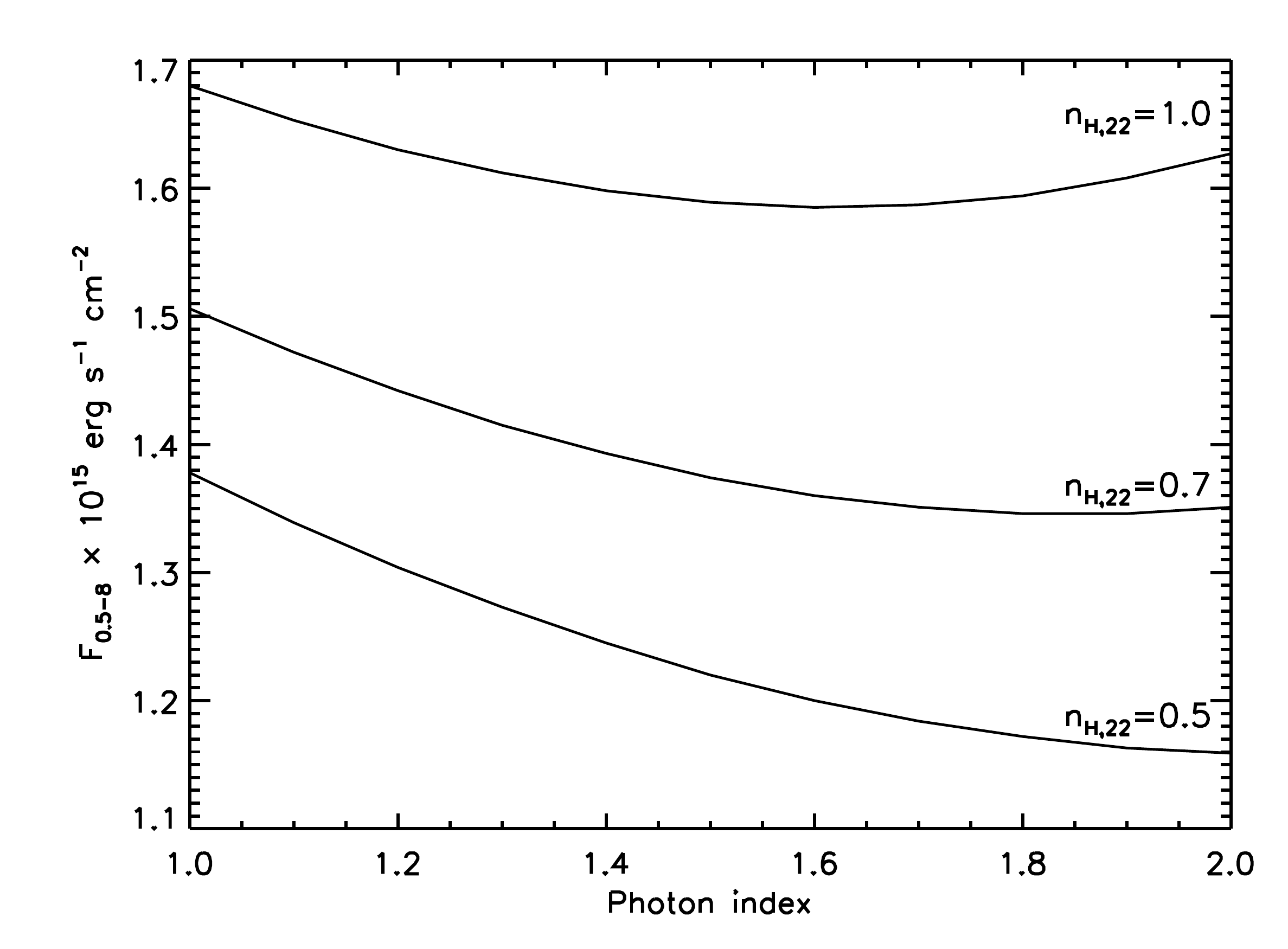}
\caption{  Upper limits on the  unabsorbed flux of J1906 in the 0.5--8 keV
band as functions of photon index for the PL spectral model,
computed
for three plausible $n_{\rm H}$ values.}
\vspace{-0.3cm}
\end{figure}

  We have also searched the ACIS image for extended X-ray emission on larger angular scales.
    The images in
   Figures 2 and 5  show a hint of  a puzzling elliptical
structure (resembling a tilted ring with a radius of $\simeq1.6'$)
   approximately centered on the J1906 position.
  To assess the statistical significance of the feature, we extracted
the surface brightness profile from a set of 12
 elliptical annuli (see Fig.\ 5).
 A $\chi^2$  fit with a  constant background (fitted to the first 7
   bins in Fig.\ 5, bottom) shows that the brightness enhancement is significant at a $3.9\sigma$ level.
    We have also extracted the ring spectrum  from the elliptical
annulus that includes
 the entire diffuse ring. The total number of extracted counts
     is 467, with the background\footnote{The background was extracted from a factor of four larger area, including a region inside the ring as well as regions
      outside the ring. The latter regions were selected on the same chip
and the same CCD nodes
 as those covered by the ring, with background point sources excluded. } contributing 81\%.
       The number of net source counts is $90\pm24 $ in the same energy range.
 After binning the spectrum heavily ($\approx90$ counts per spectral bin) and subtracting the background, we fitted an absorbed PL model (see Fig.\ 6) to the spectrum of the ring.
   For the fixed  $n_{\rm H,22}=0.7$, the spectrum fits the absorbed PL model with the photon index $\Gamma=1.6\pm0.7$ and the
   unabsorbed 0.5--8 keV flux of
$(4.0\pm0.8)\times10^{-14}$ erg s$^{-1}$ cm$^{-2}$
  ($\chi_{\nu}^2=0.98$ for $\nu=3$).

\begin{figure}
 \hspace{-0.5cm}
\includegraphics[width=2.7in,angle=90]{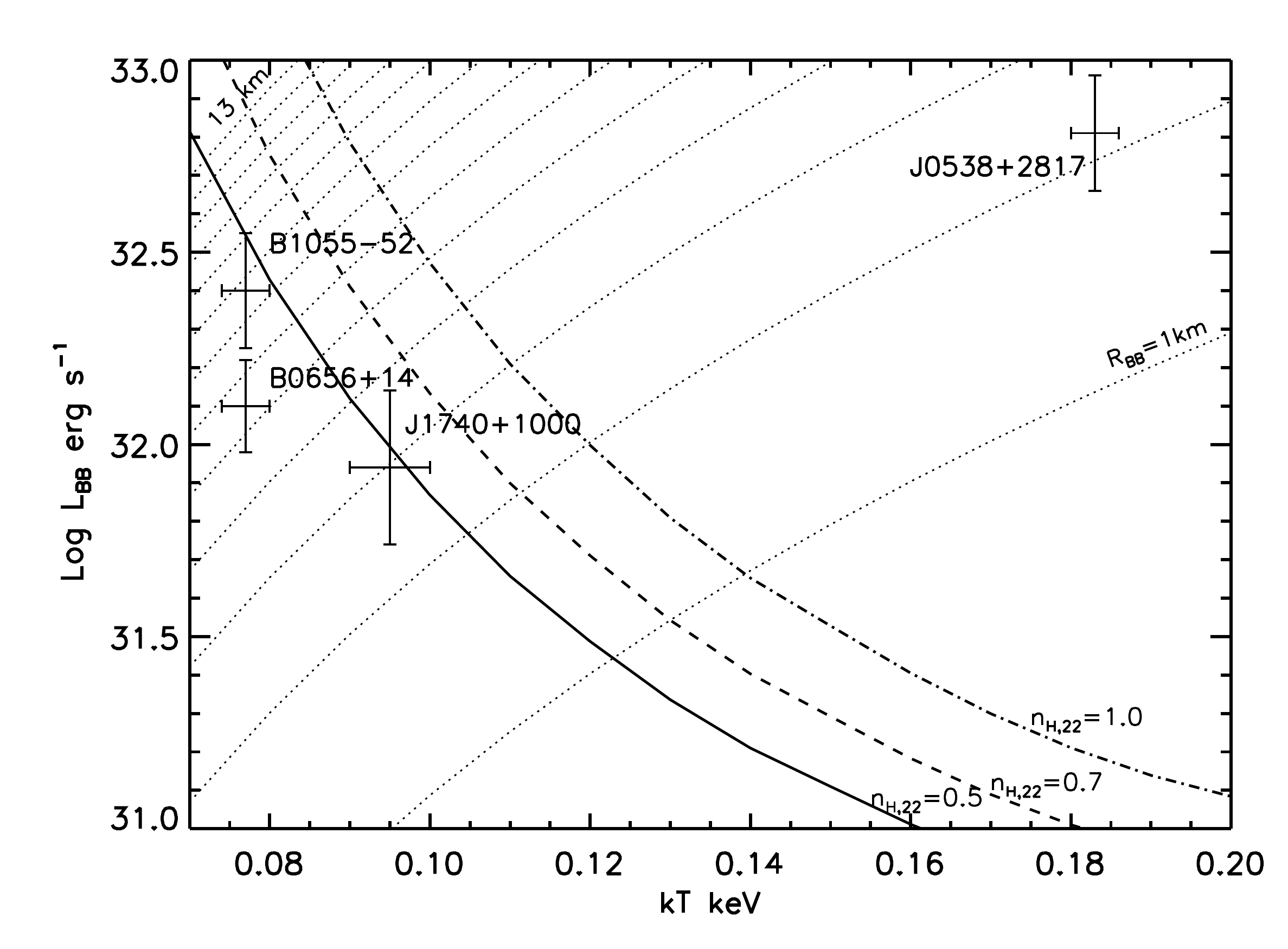}
\caption{ Upper limits on the bolometric luminosity  of thermal emission
from the surface of PSR\,J1906,
assuming a single-temperature absorbed BB model. The limits are plotted for
$n_{\rm H,22}=0.5$, 0.7, and 1,
 within a plausible range of BB temperatures as measured in pulsars with
ages similar to that of J1906. The bolometric luminosties
of these pulsars (based on  single-component BB fits; see text for details)
are
plotted for comparison. The dotted lines are loci of constant BB radii
  (increasing from lower right to upper left in 1 km increments). }
\end{figure}

\subsection{Multiwavelength Data}

  To examine the J1906 field at other wavelengths, we have searched
IR and radio survey data.
   Figures 2 and 7  show  {\sl Spitzer} IRAC images\footnote{The IRAC images
are from the GLIMPSE (Galactic Legacy
   Infrared Mid-Plane Survey Extraordinaire) data (Benjamin et al.\ 2003;  see also
   http://irsa.ipac.caltech.edu/data/SPITZER/GLIMPSE/ )} at 8 and 3.6 $\mu$m and radio images from VLA MAGPIS\footnote{See White
et al.\ (2005); also
   http://third.ucllnl.org/gps/.}.
PSR\,J1906 is
    detected in the 20 cm MAGPIS image,
 with the flux of $0.66\pm0.27$ mJy (consistent within the uncertainties
    with the $0.55\pm0.15$ mJy flux measured by
    Lorimer et al.\ 2006
in the 1.3--1.5 GHz band
from the timing observations). The measured
$1.0''\pm0.5''$ offset
   between the MAGPIC  position
     and the radio-timing position of PSR\,J1906 is consistent with both sources being the same object.

\begin{figure}
 \hspace{-0.3cm}
\includegraphics[width=3.5in,angle=0]{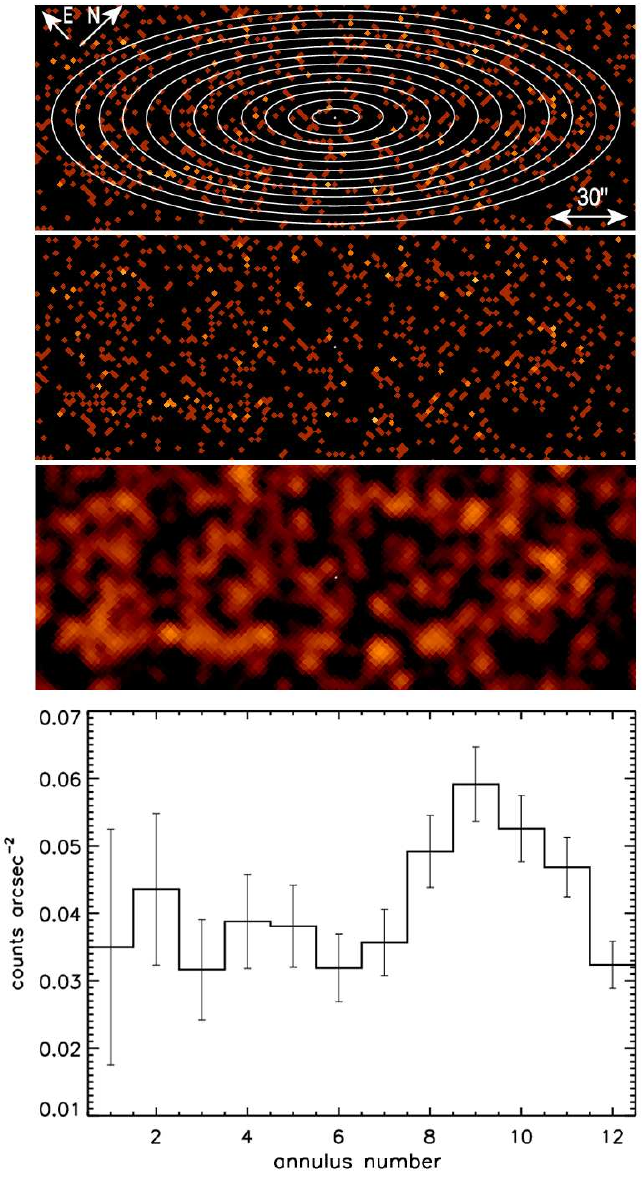}
\caption{ Three top panels show $4'\times1.5'$ ACIS images of the J1906
vicinity (in 0.5--8 keV, pixel size $1.97''$; the same part of the sky
is shown in all the images).
 The top image  shows the elliptical annuli (centered on the PSR\,J1906 radio position)
 used to characterize the ring-like structure discussed in \S3.1.
The bottom image is
smoothed with the $6''$ gaussian.  The bottom panel shows
 the average surface brightness distribution as a function of the
 annulus number
(counted outwards from the center).
}
\end{figure}

   The IR  images also
   reveal a bright ($\simeq 0.03$, 1.7, 4.3,  48.3  	and  105 Jy at 4.5, 12, 25, 60 and 100 $\mu$m, respectively\footnote{The IR fluxes are taken from
the GLIMPSE I Spring 2007 Catalog and the
IRAS Point Source Catalog (ver.\ 2.1), available at the NASA IPAC Infrared Science Archive,
    http://irsa.ipac.caltech.edu/applications/BabyGator/.}) diffuse source
IRAS~19043+0741,  $\sim46''$ west of J1906.
The source has
    a radio counterpart, GPSR5 41.594+0.160, with the flux of 4.4 mJy at 5 GHz (Becker et al.\ 1994).
    The bow-like shape of the diffuse source seen in the 8 $\mu$m image and a hint of two tails
   seen in the VLA image
   create an impression that the source is moving
    westward.  The source is  listed as a star-forming region in two catalogs (Codella et al.  1995;
   Avedisova   2002).
     No X-ray
   counterpart of this
source has been identified in the ACIS data.
 There is no reason to believe
  that J1906 was born in  IRAS~19043+0741 since
J1906
 would have to be traveling at a speed of $\sim10,000$ km s$^{-1}$
to reach its current position,
assuming the  same 5 kpc distance to the two sources and a 10 Myr
 travel time (considered to be an upper limit for an age of a star-forming region; e.g., Hartmann 2001).
   Thus, we conclude that  IRAS~19043+0741 is an unrelated object
accidently projected
near
 J0906.

 Most of the X-ray point sources seen in the images do not have
  counterparts at other wavelengths (e.g.,
the sources in Table 1 have no 2MASS counterparts), with a possible exception for the source \#3 (in Fig.\ 1 and Table 1)
located $\approx 72.8''$  northeast of
the  pulsar (i.e., outside the X-ray ring),
which apparently has a relatively bright (26 mJy at 1.4 GHz) radio counterpart NVSS~190654+074715
  but no IR counterpart.
    We did not find any traces  of the putative tilted ring (\S2.1) at any frequencies other than X-rays. A  large
number of very faint background sources (e.g.,
stars) might, in principle,  mimic the diffuse emission at least
in some parts of the putative ring.
    However,  the 3.6 $\mu$m  image does not reveal an enhanced stellar density along the ring (see Fig.\ 7) and hence does not support this interpretation.

\begin{figure}
\includegraphics[width=2.35in,angle=-90]{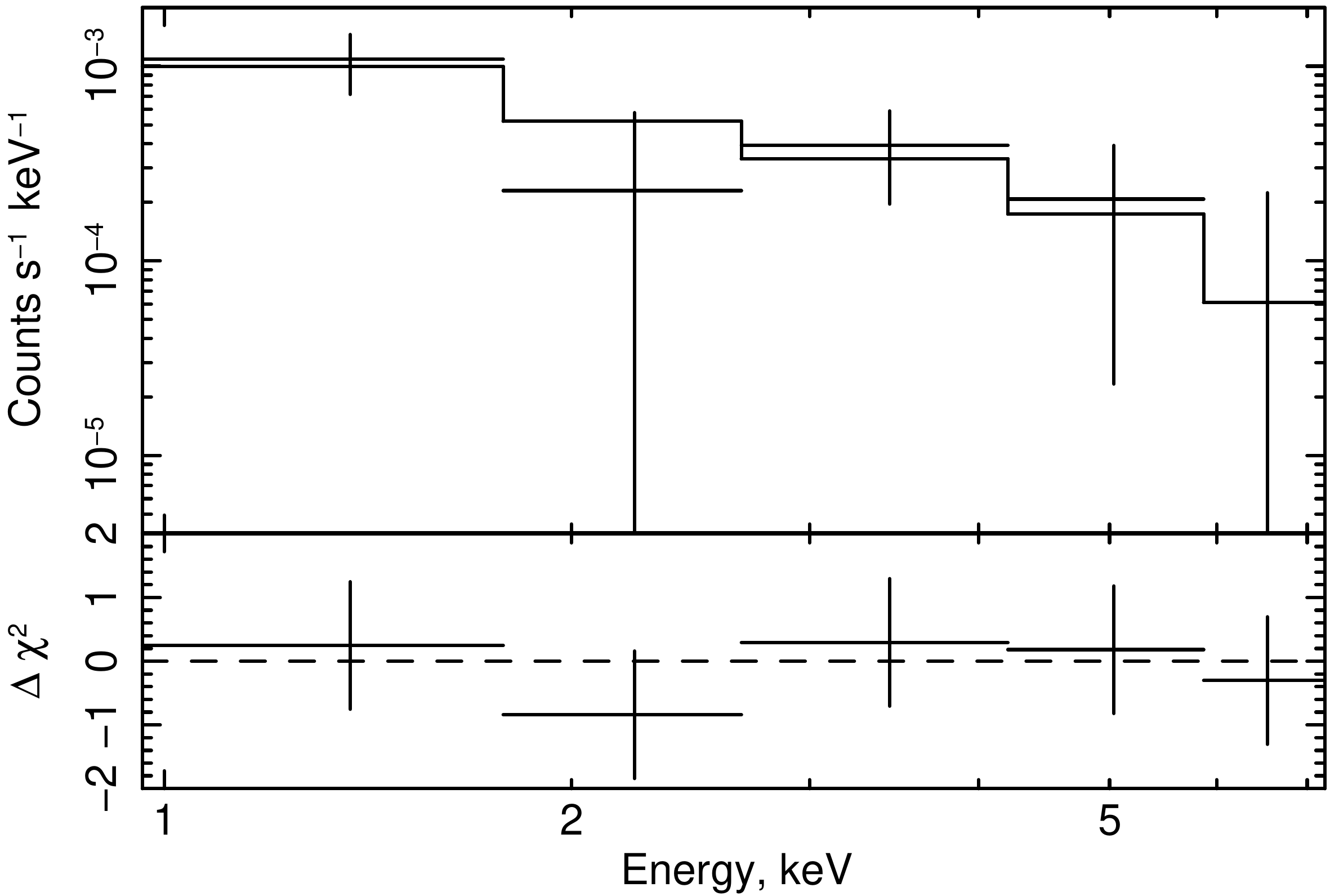}
\caption{ Spectrum of the ring (see \S2.1) fitted with the absorbed PL model ($\Gamma=1.6\pm0.7$ and fixed $n_{\rm H,22}=0.7$ ; see \S2.1). }
\end{figure}

 \begin{figure*}
 \centering
\includegraphics[width=7.1in,angle=0]{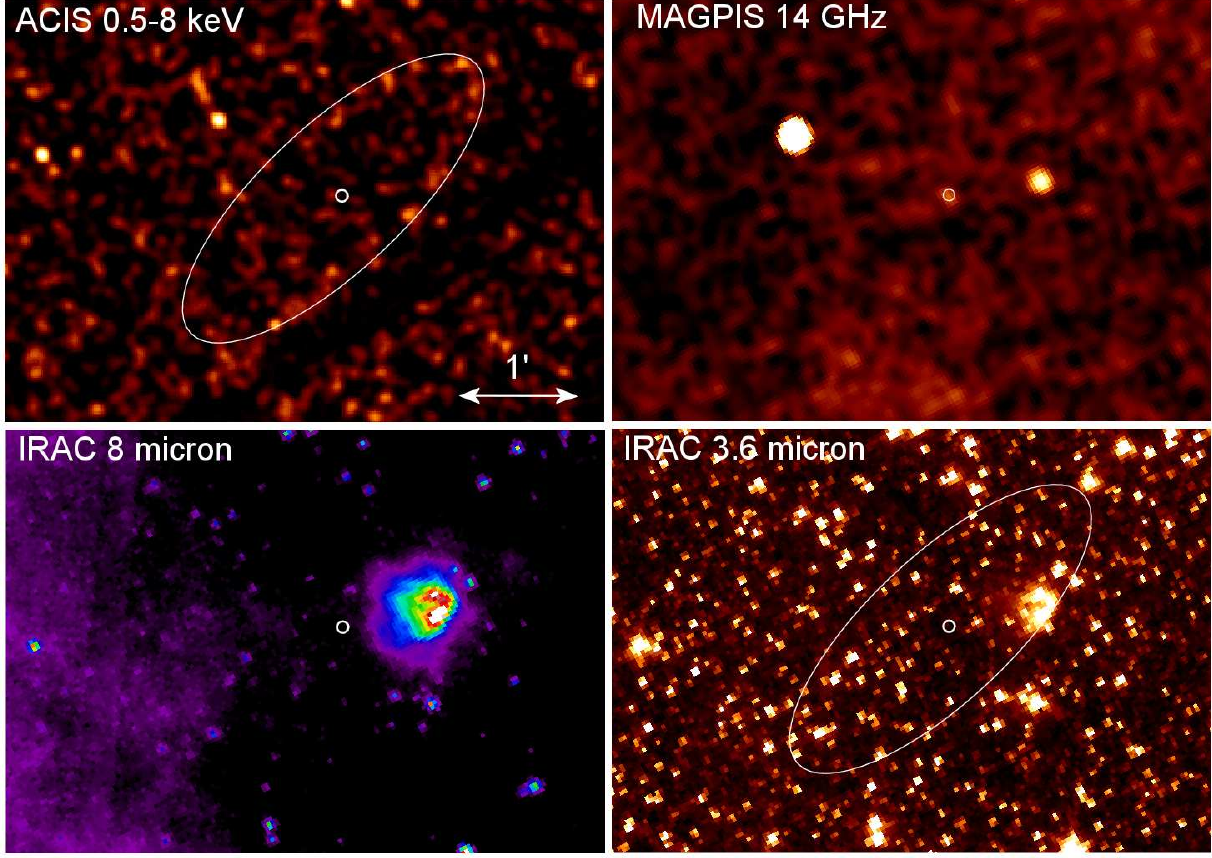}
\caption{ $3.6'\times5.0'$ images of the J1906 field
at different wavelengths (the same region of sky is shown in all the panels).
 The radio position from Lorimer et al.\ (2006) corresponds to the center
of the circle with $7''$ radius. The ellipse shows the
 region of enhanced X-ray emission shaped as an elliptical ring approximately
centered on J1906. The same ellipse in shown on the 3.6-micron image.
The ACIS image is binned to $0.98''$ pixels and smoothed with
an $r=5''$ gaussian kernel. }
\end{figure*}

\section{Discussion}

 To date,
 X-ray emission has been detected from two DNSBs (J0737--3039 and B1534+12; McLaughlin et al.\ 2004 and Kargaltsev et al.\ 2006). In both cases no extended
X-ray PWN has been seen.    However, the pulsars in these  DNSBs  are much older and less energetic than PSR\,J1906.
  The only tight binary with a pulsar (although not a DNSB)
whose PWN has been resolved in X-rays (Stappers et al.\ 2003)
is the famous ``Black Widow'' binary, in which
  the wind of the energetic  PSR B1957+20 ablates the surface of the low-mass
   dwarf companion.  Since PSR\,J1906 is even more energetic than PSR B1957+20,
 one could expect
   it to be easily detectable  in X-rays.
  Yet,
we have not detected a single photon from J1906 in the
31.6 ks {\sl Chandra} ACIS observation. We, however, discovered a puzzling diffuse emission shaped as an elliptical ring  centered on the pulsar.
 Below we discuss a possible origin of the ring, the upper limits on the J1906 X-ray luminosity, and
 the constraints on the physical properties
 of this unique system.

  PSR\,J1906 is sufficiently young and energetic to power a detectable X-ray PWN.
   One  can also expect to see  thermal X-ray emission from the hot surface of the 100-kyr-old NS and possibly non-thermal emission from the pulsar's magnetosphere.
In addition,  the
presence of a NS (or heavy WD) companion might perturb the PSR\,J1906 wind
 and  give rise to an additional   emission component
from an intrabinary shock.
  However, in J1906 the latter  component
   should  be much fainter compared to the other emission components mentioned above.
    Indeed, if the J1906 companion is a WD,
 then the
solid angle subtended by the WD,
$\Delta\Omega\simeq (R_{\rm WD}/2 a)^2 =
1.3\times 10^{-6} (R_{\rm WD}/2\times 10^8\,{\rm cm})^2$ sr,
is so
small that the WD cannot intercept a significant fraction of the
pulsar's wind, even if the wind is strongly anisotropic.
The fraction
of the intercepted wind could be larger if the companion is a NS
because in this case the effective cross section  would be determined
by the standoff distance,
at which the magnetic pressure of the companion's  magnetosphere is equal
to the pressure of the
pulsar's wind
(e.g., Kargaltsev et al.\ 2006). It should be noted, however,  that the
  companion NS is likely significantly older than PSR\,J1906
 because it  takes at least $10^{6}$--$10^{7}$ years to spin up
this first-formed NS and expel the outer layers of its companion
(progenitor of PSR\,J1906)
through the common envelope phase,
  thereby shrinking the binary orbit  to its current size.
In addition,
the first-born NS would only retain
a weak magnetic field, $\lesssim10^{10}$ G, after a long period of accretion.
Thus, if the J1906's companion NS has properties similar to those of
PSR J0737--3039A,   then the size of the compressed
  companion magnetosphere,
$\sim 7\times10^{8}$ cm
(see eq.\ [1] in Arons et al.\ 2005), is comparable to a typical WD radius,
i.e., only a small fraction of the J1906's wind would be intercepted.
    Thus,  one   would expect  the observed X-ray properties of the J1906  PWN to be
   such as  what they would be if the PWN were powered by a solitary pulsar.

 Finally, one could expect to see  X-ray emission from the companion if it is
indeed a NS. However, according to  the evolutionary scenarios (see above)
the companion NS is expected to be
 much older, with a colder surface and lower $\dot{E}$.
Hence, its X-ray emission is expected to be significantly fainter than
that of PSR\,J1906.

\subsection{Limits on the nonthermal X-ray emission from the  pulsar and unresolved compact PWN.}

     The upper limit on the combined unabsorbed   flux from the compact, unresolved PWN and the pulsar's
      magnetosphere, estimated  as described in \S2.1,
  is $(1.1-1.7)\times10^{-15}$ erg s$^{-1}$ cm$^{-2}$ at 90\% confidence,
in the 0.5--8 keV band
      (see Fig.\ 3).
      This flux corresponds to the luminosity
$L_X<(3.3-5.1)\times10^{30}d_{5}^2$ erg s$^{-1}$ and
       efficiency
  $\eta_X <
(1.2-1.9)\times 10^{-5}d_{5}^2$.
Compared to other pulsar + PWN systems, the inferred  upper limits
are unusually low, albeit not the most extreme.
This is seen from Figure 8, in which we plotted the 0.5--8 keV pulsar + PWN
luminosities for about 40 detected systems, and a few upper limits,
versus pulsar's spin-down power $\dot{E}$ (see also KP08).
In particular, the pulsar + PWN
luminosities and efficiencies for the objects with similar $\dot{E}$ values
are a factor of a few higher than the J1906's upper limits
(see Table 2).
Yet, there is a younger, more energetic pulsar, J1913+1011,
from which no counts were detected in a 20 ks {\sl Chandra} ACIS-S3 observation,
and for which the upper limit on the pulsar + PWN efficiency is even lower
than for J1906 (see Fig.\ 8). The nondetection of pulsars could be attributed
to an unfavorable orientation of the X-ray beams, but it is more difficult to
explain the nondetction of PWNe, whose emission should be more
isotropic (and which
are, on average,  a factor of $\sim$4 more luminous in X-rays than their
pulsars; Kargaltsev et al.\ 2007).

Thus,  although
the X-ray efficiency of the J1906 PWN  is  within the range of efficiencies
  measured  for the entire sample of  pulsar + PWN sources,
  J1906 is
among the
most underluminous ones.
    We should stress that the observed {\em four orders of magnitude}
scatter in $\eta_X$
cannot be
   attributed just to poorly measured distances,
and therefore  better understanding of pulsar wind physics  is needed
    to explain the very different  X-ray pulsar + PWN efficiencies.

\subsection{
Interpretation of the  ring: An unusual PWN?}

 The nature of the tentatively detected ring-like structure (see \S2.1)
is puzzling.
One might assume that the ring is a limb-brightened shell of the
remnant of the supernova in which PSR\,J1906 was born. However, the
size of the shell is too small for a SNR of a 100 kyr age. For instance,
in the Sedov regime, the SNR radius can be estimated as
$\simeq 21' (E_{51}/n)^{1/5}(t/100~{\rm kyrs})^{2/5}d_5^{-1}$ (where
$E=10^{51}E_{51}$ erg s$^{-1}$ is the SN energy release, and $n$ is the
ambient matter density in units of cm$^{-3}$),
much larger than
the $1.6'$ semimajor axis of the observed ellipse, at reasonable values
of the parameters. To explain the
ellipsoidal shape of the shell, an unusual density distribution in the
ambient medium would be required. Moreover, it is hard to expect
such a narrow, distinct shell for so old SNR. Finally,
the spectrum of the elliptical ring ($\Gamma =1.6\pm 0.7$,
detected up to 7 keV, albeit with a scant statistic) is too hard for
an SNR shell (its slope is typical for a PWN spectrum, however; KP08).
Therefore, we do not consider the SNR shell interpretation to be plausible.

 Alternatively,
one could attempt to
interpret the ring as synchrotron emission produced
  immediately
  downstream of the termination shock in the pulsar wind.
Although such structures
  have been
seen in many PWNe (e.g.,
the rings around the Crab and J1930+1852 pulsars),
the X-ray emission from the
   pulsar (or its unresolved vicinity)
   is
usually more luminous than that of the ring (see Fig.\ 2 in KP08).
The only
apparent exception is the PWN in the IC\,443 SNR
(Gaensler et al. 2006; Weisskopf et al.\ 2007),
where the alleged pulsar (pulsations are yet to
be found) is a factor of 10 dimmer (in terms
of the 0.5--8 keV flux) than the surrounding bright PWN emission
associated with the termination shock
(which, however, is not clearly resolved into a ring in this case).
 The nondetection of PSR\,J1906
      implies that it is at least a factor of 20--40  dimmer than the ring.
  The reasons for such a low pulsar-to-PWN flux ratio
    could be
   a special orientation
   of  the magnetic and rotation axes resulting in suppressed magnetospheric activity,  or
     the presence of absorbing matter near the pulsar (Cordes \& Shannon 2008).
    In addition to the problematic flux ratio, the observed size of the ring requires stretching the limits on other physical parameters.
   Since most of the PSR\,J1906's wind escapes unperturbed by the interaction with the companion (see above),
     the termination shock in the pulsar wind
   would occur at $R_{\rm TS} \approx (\dot{E}f_w/4\pi c P_{\rm amb})^{1/2} =
 0.085 f_w^{1/2}P_{\rm amb,-11}^{-1/2}$ pc (it corresponds to the
angular distance of $3.5'' f_w^{1/2}P_{\rm amb,-11}^{-1/2}d_5^{-1}$),
where $P_{\rm amb}\equiv 10^{-11} P_{\rm amb,-11}$ dyne cm$^{-2}$ is the
ambient pressure, and the factor $f_w$ takes into account possible
anisotropy of the pulsar wind (e.g., Kargaltsev et al.\ 2006) .
This distance is much smaller than the $1.6'$ radius of the
putative tilted ring,
unless the pulsar wind is extremely well collimated
in the equatorial plane and/or the ambient  pressure is very low
(e.g., $f_w \sim 100$, $P_{\rm amb,-11}\sim0.1$).
These requirements can, however, be
  relaxed if the pulsar is substantially
closer than 5 kpc
(measuring the pulsar's parallax
is needed to establish the distance reliably).
On the other hand, the unabsorbed luminosity of the ring,
$L_X=(1.2\pm0.3)\times10^{32}d_5^2$ erg s$^{-1}$
for the plausible $n_{\rm H,22}=0.7$, gives the X-ray efficiency
$\eta_X
=4.5\times10^{-4}d_5^2$, typical for
 a PWN (see the dashed errorbar in Fig.\ 8).

Another
conceivable interpretation of the observed structure might be a ``light echo''
 caused by reflection from  dust
   following  a hypothetical flare   (e.g., due to an accretion episode)
   that could occur in the binary  in the recent past (e.g., a few   years ago).  We know nothing about the
    putative  flare, nor about the dust distribution around the system.
   However,    the elliptical shape of the structure would require a very special  dust distribution.
   In this scenario, which currently seems rather unlikely,
 a subsequent observation might show an increased size of the structure.

 Finally, there still remains a possibility that  the ring-like structure  might be caused by  fluctuations in the detector background (which could be
enhanced on one side because of
the proximity to the chip edge\footnote{
However, we do not see an enhanced surface brightness in
 the regions adjacent to the boundaries of the other I-array chips.})
and sky background (e.g., due to a number of faint, unresolved
point sources, such as field stars, accidentally projected along
the ellipse).
A deeper {\sl Chandra} observation, with the target placed farther from the chip gaps, can
    ascertain  the reality of the enigmatic ring.

\begin{figure}
\hspace{-0.5cm}
\includegraphics[width=2.8in,angle=90]{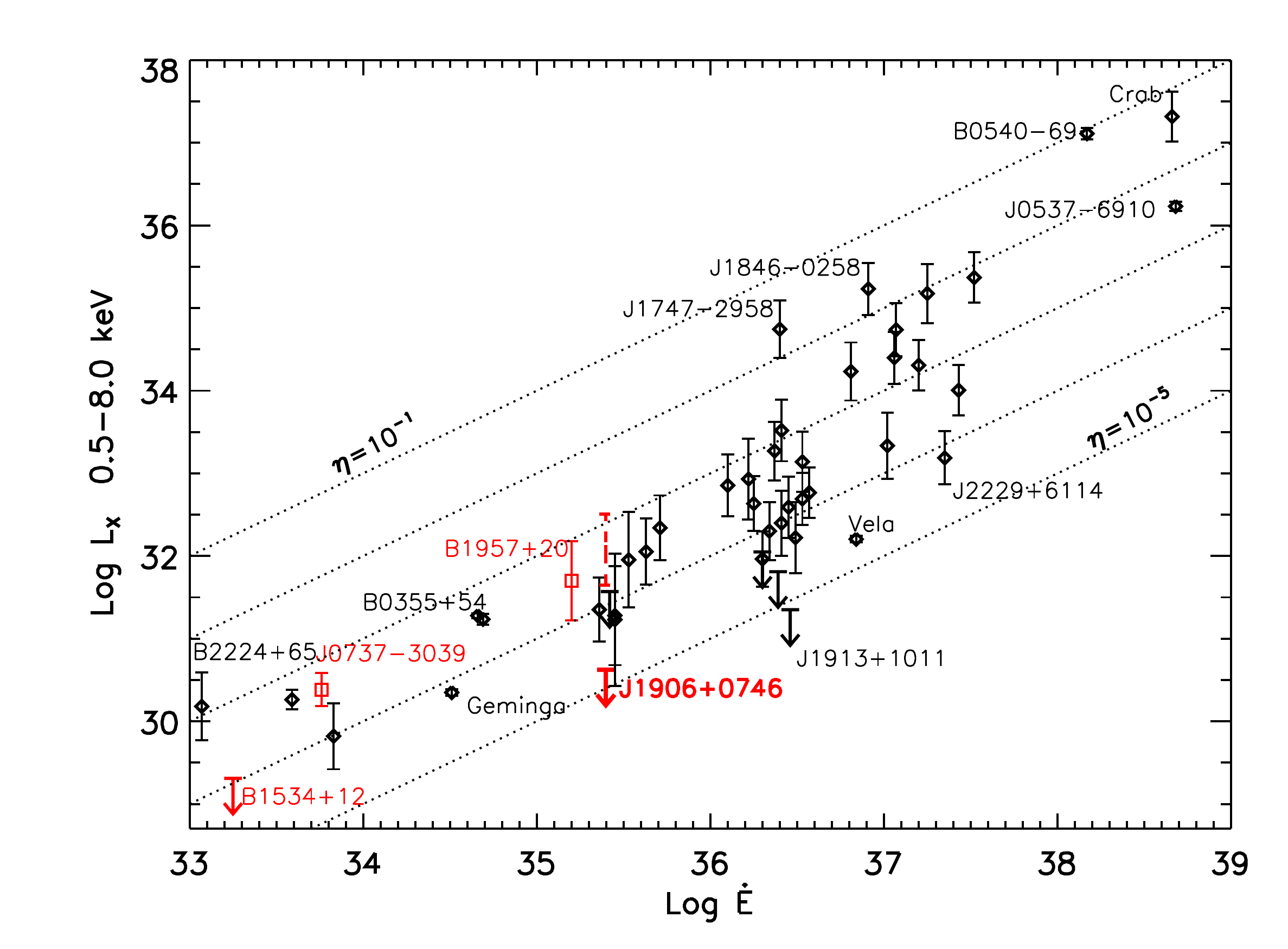}
\caption{ Summed nonthermal 0.5--8 keV
luminosities of pulsar + PWN systems
 observed with {\sl Chandra} versus pulsar's spin-down power.
The tight pulsar binaries discussed in the paper are shown in red.
 The dashed red errorbar corresponds to the luminosity of the elliptical ring
which
might be an extended PWN associated with J1906. }
\end{figure}

\subsection{Limits on thermal emission from the  NS surface}

The surface of a 100-kyr-old NS is expected to be hot enough to emit
soft X-rays (Yakovlev \& Pethick 2004, and references therein).
The surface temperature is likely non-uniform,
  with polar regions being hotter than the rest of the NS surface
(see, e.g., Pavlov et al.\ 2002; De Luca et al.\ 2005).
The observation of J1906 allows us to put an upper limit on the bolometric
luminosity, $L_{\rm BB}=4\pi R^{2}\sigma_{\rm SB} T^{4}$, of PSR\,J1906
and compare it with the luminosities of other pulsars of similar ages.
 Figure 4 shows the upper limits in the $T$-$L_{\rm BB}$ plane,
computed  in a plausible range of
temperatures for three $n_{\rm H}$ values, assuming a BB spectrum.
The same figure shows the BB temperatures and bolometric luminosities
 for four other pulsars with similar spin-down ages\footnote{
The parameters of the BB fits are from our own analysis, except for J0538+2817 (Zavlin \& Pavlov 2004).}.
We see from Figure 4 that the upper limit curves $L_{\rm BB}(T)$ for
PSR\,J1906 ($\tau=112$ kyr) pass just slightly above the
points corresponding to
PSR J1740+1000
($\tau=115$ kyr),
B0656+14
($\tau=110$ kyr),
  and B1055--52
($\tau=537$ kyr).
On the contrary, the point corresponding to
PSR J0538+2817
($\tau=617$ kyrs) lie
substantially above the upper limit curves because of the high temperature
(and small BB radius) of this pulsar.
 However, based on the measurements of the proper motion and parallax
of PSR J0538+2817, which resides in the S147 SNR, its true age, 20--60 kyr,
is much smaller than the spin-down age\footnote{
The apparent discrepancy between the spin-down and kinematic ages
 can be eliminated if we assume that the pulsar was born with a period
close to its modern value (Ng et al.\ 2007).},
which may explain the high temperature (Ng et al.\ 2007).
    From the upper limits  in Figure 4 we can conclude that PSR\,J1906 is unlikely to be significantly  younger  than  its
spin-down age,
and the BB luminosity
    of PSR\,J1906
may be similar to those of the other three pulsars
[$L_{\rm BB}\approx(1-
3)\times10^{32}$ erg s$^{-1}$].

We should note that the single-component BB model, used above for the
sake of comparison, provides only a crude approximation to the
spectra of the thermally-emitting pulsars. For instance,
the high-quality {\sl XMM-Newton} spectrum of PSR J1740+1000,
 which is almost  an exact twin
of J1906 in terms of the spin-down properties,
is much better described
by  a  three-component model, with two BB components
     ($kT\approx70$ and $140$ eV, and $R\approx5.8$ and $0.6$ km,
respectively) and a PL component with  $\Gamma=1.2\pm0.3$
(Misanovic et al., in preparation).
However, the bolometric luminosity of the low-temperature
      component, $\simeq 1\times 10^{32}$ erg s$^{-1}$, is close to that
inferred from the simplified  BB description of the PSR J1740+1000's
X-ray spectrum, thus
supporting the simplified analysis described above.

\section{Summary}

 In the
31.6 ks
{\sl Chandra }  ACIS observation we have not detected a single photon
from the J1906 binary within the $3''$ radius from its radio position.
The nondetection of PSR\,J1906 and its compact PWN
could be explained by several
factors.
The thermal X-ray emission from the surface of the middle-aged PSR\,J1906 is
likely too strongly absorbed by the ISM to detect it in a short observation,
but the upper limit on the bolometric BB luminosity only slightly exceeds
the bolometric luminosities observed in pulsars of similar ages.
The nondetection of the
pulsar's magnetospheric emission might be explained by an unfavorable
orientation of the pulsar beam.
The lack of X-ray emission from an intrabinary shock
can be explained by a small size of the intrabinary interaction region
that   intercepts only a tiny fraction of the PSR\,J1906 wind.
However,
the non-detection of a compact PWN around the J1906 binary is surprising;
perhaps it could be ascribed to unusual properties of the ambient medium
or of the pulsar itself.

    We found tentative
evidence of a puzzling elliptical ring centered on J1906,
with a $3.9\sigma$ significance; however,
   the  size of the structure  is significantly larger than the
stand-off distance to the termination shock produced by an
 isotropic pulsar wind in a typical ISM.
    The PWN interpretation of the ring is not ruled out,
but it requires a  high degree of
the pulsar wind collimation and a low density of the surrounding ISM,
 or a distance significantly smaller than estimated from the pulsar's
dispersion measure.
   Deeper, high-resolution observations at X-ray and radio frequencies
should help to clarify the nature of the observed diffuse structure and
to unambiguously
   detect the elusive pulsar wind from this unique pulsar.

\acknowledgements Support for this work was provided by the
National Aeronautics and Space Administration through Chandra
Award Number G07-8061X
 issued by the Chandra X-ray Observatory Center,
which is operated by the Smithsonian Astrophysical Observatory for
and on behalf of the National Aeronautics Space Administration
under contract NAS8-03060. The work was also partially supported by
NASA grant NNX09AC84G.

\begin{table*}[]
 \caption{Properties  of pulsars with spin-down parameters  similar to PSR\,J1906.}
 \begin{center}
 \setlength{\tabcolsep}{0.1in}
\begin{tabular}{lrcccccccccc}
\tableline\tableline
   PSR &    $P$  &  $\log\tau$ &   $\log\dot{E}$ &  $d$\tablenotemark{a} &
    $n_{\rm H,22}$\tablenotemark{b} & Exp.\tablenotemark{c} &
      $N_{\rm cts}$\tablenotemark{d} &
       $\log L_{\rm PSR}$\tablenotemark{e}  &
        $\log L_{\rm PWN}$\tablenotemark{f} &
        $\log \eta_{\rm PSR}$\tablenotemark{g} &
         $\log \eta_{\rm PWN}$\tablenotemark{h}
   \\
  \tableline
  & ms &     & & kpc &  & ks&  cts & &    \\
\tableline
 J1906+0746    &  144  &  5.05  &  35.43  & 5     &
0.7 &  31.6    & 0         & $<30.7$ & $<30.7$[32.1] & $< -4.7$ & $<-4.7$ [$-3.3$] \\
 J1702--4128    &  182   &  4.74  &  35.53  & 5    &  1.1  &  10.4    & 8         & $31.70\pm0.50$ & $31.60\pm0.50$    &  $-3.8$ &$-3.9$  \\
 J0729--1448    &  252   &  4.54  &  35.45  & 4    & 0.3   &  4.1      & 13       & $31.30\pm0.30$ &   $ 31.20\pm0.50$ & $-4.1$ & $-4.2$ \\
J1740+1000    &  154   &  5.06  &  35.36  & 1.4  & 0.1   &  5.1      & 130     & $31.10\pm0.05$ & $ 31.11\pm0.10$   &$-4.3$ & $-4.3$  \\
J1841--0345    &   204  &  4.75 &  35.43   & 4      &  0.8  &   10.0   &  2       & $<31.57$   & $<31.57$   & $< -3.9$ & $< -3.9$  \\
\tableline
\end{tabular}
\end{center}
\tablenotetext{a}{
Distance estimate based on the pulsar's dispersion
measure.}
\tablenotetext{b}{Hydrogen column density
estimated  from the pulsar's dispersion measure (assuming 10\%
ISM ionization).}
\tablenotetext{c}{{\sl Chandra} ACIS exposure
time.}
\tablenotetext{d}{Number of counts  in the 0.5--8 keV band,
calculated for an  $R=3''$ circular aperture centered on the radio
pulsar position.}
\tablenotetext{e}{~Logarithm of the pulsar
luminosity (or 90\% upper limit on the pulsar + compact PWN luminosity
for undetected objects) in the
0.5--8 keV band.
For J1740+1000, the quoted luminosity is
for the PL component only; the total 0.5-8 keV luminosity,  $10^{32.05}$
erg s$^{-1}$, is much higher because of the contribution of the thermal
component. }
\tablenotetext{f}{~Logarithm of the PWN
luminosity (or 90\% upper limit on pulsar + compact PWN luminosity
for undetected objects) in the 0.5--8 keV band. The luminosity for
the J17400+1000 PWN does not include the very extended tail
(Kargaltsev et al.\ 2008).
For J1906, we also provide the luminosity
of the putative ring in the brackets.}
\tablenotetext{g}{~Logarithm of the pulsar
 X-ray efficiency (or 90\% upper limit).}
\tablenotetext{h}{~Logarithm of the PWN
 X-ray efficiency (or 90\% upper limit).}
\end{table*}

\end{document}